\RequirePackage{lineno}

\documentclass[aps,prl,twocolumn,showpacs,superscriptaddress,groupedaddress]{revtex4}  
\usepackage{graphicx}  
\usepackage{dcolumn}   
\usepackage{bm}        
\usepackage{amssymb}   

\usepackage{nicefrac}
\usepackage{enumerate}

\def\jpsi{$J/\psi$}                     
\def\ppbar{$p\overline{p}$}             
\def\B0s{$B_s^0$}			 
\def\aB0s{$\overline{B_s^0}$}            
\def\Lb{$\Lambda_b^0$}			 
\def\L0{$\Lambda^0$}			 
\def\La0{$\Lambda^0$}			 
\def\aL0{$\overline{\Lambda^0}$}      
\def\Lbch{$\Lambda_b^0 \rightarrow J/\psi \, \Lambda^0$}  
\def\Bch{$B^0 \rightarrow J/\psi \, K_S^0$}  
\def\B0{$B^0$}			 
\def\aB0{$\overline{B^0}$}      
\def\K0{$K_S^0$}			 
\def\V0{$V^0$}			 
\def\ifb{fb$^{-1}$}                     
\def\gevcc{GeV/$c^2$}                   
\def\gevc{GeV/$c$}                         
\def\D0{D0}                            

\begin{document}


\hspace{5.2in} \mbox{FERMILAB-PUB-12-097-E}

\title{Measurement of the $\boldsymbol{\Lambda_b^0}$~lifetime in the exclusive decay $\boldsymbol{\Lambda_b^0 \rightarrow J/\psi \Lambda^0}$~in $\boldsymbol{p\bar{p}}$~collisions at $\boldsymbol{\sqrt{s}=1.96}$~TeV}
\affiliation{LAFEX, Centro Brasileiro de Pesquisas F\'{i}sicas, Rio de Janeiro, Brazil}
\affiliation{Universidade do Estado do Rio de Janeiro, Rio de Janeiro, Brazil}
\affiliation{Universidade Federal do ABC, Santo Andr\'e, Brazil}
\affiliation{University of Science and Technology of China, Hefei, People's Republic of China}
\affiliation{Universidad de los Andes, Bogot\'a, Colombia}
\affiliation{Charles University, Faculty of Mathematics and Physics, Center for Particle Physics, Prague, Czech Republic}
\affiliation{Czech Technical University in Prague, Prague, Czech Republic}
\affiliation{Center for Particle Physics, Institute of Physics, Academy of Sciences of the Czech Republic, Prague, Czech Republic}
\affiliation{Universidad San Francisco de Quito, Quito, Ecuador}
\affiliation{LPC, Universit\'e Blaise Pascal, CNRS/IN2P3, Clermont, France}
\affiliation{LPSC, Universit\'e Joseph Fourier Grenoble 1, CNRS/IN2P3, Institut National Polytechnique de Grenoble, Grenoble, France}
\affiliation{CPPM, Aix-Marseille Universit\'e, CNRS/IN2P3, Marseille, France}
\affiliation{LAL, Universit\'e Paris-Sud, CNRS/IN2P3, Orsay, France}
\affiliation{LPNHE, Universit\'es Paris VI and VII, CNRS/IN2P3, Paris, France}
\affiliation{CEA, Irfu, SPP, Saclay, France}
\affiliation{IPHC, Universit\'e de Strasbourg, CNRS/IN2P3, Strasbourg, France}
\affiliation{IPNL, Universit\'e Lyon 1, CNRS/IN2P3, Villeurbanne, France and Universit\'e de Lyon, Lyon, France}
\affiliation{III. Physikalisches Institut A, RWTH Aachen University, Aachen, Germany}
\affiliation{Physikalisches Institut, Universit\"at Freiburg, Freiburg, Germany}
\affiliation{II. Physikalisches Institut, Georg-August-Universit\"at G\"ottingen, G\"ottingen, Germany}
\affiliation{Institut f\"ur Physik, Universit\"at Mainz, Mainz, Germany}
\affiliation{Ludwig-Maximilians-Universit\"at M\"unchen, M\"unchen, Germany}
\affiliation{Fachbereich Physik, Bergische Universit\"at Wuppertal, Wuppertal, Germany}
\affiliation{Panjab University, Chandigarh, India}
\affiliation{Delhi University, Delhi, India}
\affiliation{Tata Institute of Fundamental Research, Mumbai, India}
\affiliation{University College Dublin, Dublin, Ireland}
\affiliation{Korea Detector Laboratory, Korea University, Seoul, Korea}
\affiliation{CINVESTAV, Mexico City, Mexico}
\affiliation{Nikhef, Science Park, Amsterdam, the Netherlands}
\affiliation{Radboud University Nijmegen, Nijmegen, the Netherlands}
\affiliation{Joint Institute for Nuclear Research, Dubna, Russia}
\affiliation{Institute for Theoretical and Experimental Physics, Moscow, Russia}
\affiliation{Moscow State University, Moscow, Russia}
\affiliation{Institute for High Energy Physics, Protvino, Russia}
\affiliation{Petersburg Nuclear Physics Institute, St. Petersburg, Russia}
\affiliation{Instituci\'{o} Catalana de Recerca i Estudis Avan\c{c}ats (ICREA) and Institut de F\'{i}sica d'Altes Energies (IFAE), Barcelona, Spain}
\affiliation{Uppsala University, Uppsala, Sweden}
\affiliation{Lancaster University, Lancaster LA1 4YB, United Kingdom}
\affiliation{Imperial College London, London SW7 2AZ, United Kingdom}
\affiliation{The University of Manchester, Manchester M13 9PL, United Kingdom}
\affiliation{University of Arizona, Tucson, Arizona 85721, USA}
\affiliation{University of California Riverside, Riverside, California 92521, USA}
\affiliation{Florida State University, Tallahassee, Florida 32306, USA}
\affiliation{Fermi National Accelerator Laboratory, Batavia, Illinois 60510, USA}
\affiliation{University of Illinois at Chicago, Chicago, Illinois 60607, USA}
\affiliation{Northern Illinois University, DeKalb, Illinois 60115, USA}
\affiliation{Northwestern University, Evanston, Illinois 60208, USA}
\affiliation{Indiana University, Bloomington, Indiana 47405, USA}
\affiliation{Purdue University Calumet, Hammond, Indiana 46323, USA}
\affiliation{University of Notre Dame, Notre Dame, Indiana 46556, USA}
\affiliation{Iowa State University, Ames, Iowa 50011, USA}
\affiliation{University of Kansas, Lawrence, Kansas 66045, USA}
\affiliation{Kansas State University, Manhattan, Kansas 66506, USA}
\affiliation{Louisiana Tech University, Ruston, Louisiana 71272, USA}
\affiliation{Boston University, Boston, Massachusetts 02215, USA}
\affiliation{Northeastern University, Boston, Massachusetts 02115, USA}
\affiliation{University of Michigan, Ann Arbor, Michigan 48109, USA}
\affiliation{Michigan State University, East Lansing, Michigan 48824, USA}
\affiliation{University of Mississippi, University, Mississippi 38677, USA}
\affiliation{University of Nebraska, Lincoln, Nebraska 68588, USA}
\affiliation{Rutgers University, Piscataway, New Jersey 08855, USA}
\affiliation{Princeton University, Princeton, New Jersey 08544, USA}
\affiliation{State University of New York, Buffalo, New York 14260, USA}
\affiliation{Columbia University, New York, New York 10027, USA}
\affiliation{University of Rochester, Rochester, New York 14627, USA}
\affiliation{State University of New York, Stony Brook, New York 11794, USA}
\affiliation{Brookhaven National Laboratory, Upton, New York 11973, USA}
\affiliation{Langston University, Langston, Oklahoma 73050, USA}
\affiliation{University of Oklahoma, Norman, Oklahoma 73019, USA}
\affiliation{Oklahoma State University, Stillwater, Oklahoma 74078, USA}
\affiliation{Brown University, Providence, Rhode Island 02912, USA}
\affiliation{University of Texas, Arlington, Texas 76019, USA}
\affiliation{Southern Methodist University, Dallas, Texas 75275, USA}
\affiliation{Rice University, Houston, Texas 77005, USA}
\affiliation{University of Virginia, Charlottesville, Virginia 22901, USA}
\affiliation{University of Washington, Seattle, Washington 98195, USA}
\author{V.M.~Abazov} \affiliation{Joint Institute for Nuclear Research, Dubna, Russia}
\author{B.~Abbott} \affiliation{University of Oklahoma, Norman, Oklahoma 73019, USA}
\author{B.S.~Acharya} \affiliation{Tata Institute of Fundamental Research, Mumbai, India}
\author{M.~Adams} \affiliation{University of Illinois at Chicago, Chicago, Illinois 60607, USA}
\author{T.~Adams} \affiliation{Florida State University, Tallahassee, Florida 32306, USA}
\author{G.D.~Alexeev} \affiliation{Joint Institute for Nuclear Research, Dubna, Russia}
\author{G.~Alkhazov} \affiliation{Petersburg Nuclear Physics Institute, St. Petersburg, Russia}
\author{A.~Alton$^{a}$} \affiliation{University of Michigan, Ann Arbor, Michigan 48109, USA}
\author{G.~Alverson} \affiliation{Northeastern University, Boston, Massachusetts 02115, USA}
\author{M.~Aoki} \affiliation{Fermi National Accelerator Laboratory, Batavia, Illinois 60510, USA}
% ...
%
%
\collaboration{The D0 Collaboration\footnote{with visitors from
$^{a}$Augustana College, Sioux Falls, SD, USA,
$^{b}$The University of Liverpool, Liverpool, UK,
$^{c}$UPIITA-IPN, Mexico City, Mexico,
$^{d}$DESY, Hamburg, Germany,
,
$^{e}$SLAC, Menlo Park, CA, USA,
$^{f}$University College London, London, UK,
$^{g}$Centro de Investigacion en Computacion - IPN, Mexico City, Mexico,
$^{h}$ECFM, Universidad Autonoma de Sinaloa, Culiac\'an, Mexico
and
$^{i}$Universidade Estadual Paulista, S\~ao Paulo, Brazil.
$^{\ddag}$Deceased.
}} \noaffiliation
\vskip 0.25cm
    
\date{April 12, 2012}

\begin{abstract}
We measure the \Lb~lifetime  in the  fully reconstructed decay  \Lbch~using 10.4 \ifb~of \ppbar~collisions collected with the \D0~detector at $\sqrt{s}=1.96$ TeV. The lifetime of the topologically similar decay channel \Bch~is also measured.
We obtain 
$\tau(\Lambda_b^0) =   1.303 \pm 0.075 \text{ (stat.)} \pm 0.035 \text{ (syst.)}$~ps 
and  
$\tau(B^0) =   1.508 \pm 0.025 \text{ (stat.)} \pm 0.043 \text{ (syst.)}$~ps. 
Using these measurements, we determine the lifetime ratio of  
$\tau(\Lambda_b^0)/\tau(B ^0) = 0.864 \pm 0.052 \text{ (stat.)} \pm 0.033 \text{ (syst.)} $.
\end{abstract}

\maketitle

Lifetime measurements of particles containing $b$ quarks provide important tests of the significance of strong interactions between the constituent partons in the weak decay of $b$ hadrons.
These interactions produce measurable differences between $b$ hadron lifetimes that the heavy quark expansion (HQE)~\cite{HQE} predicts with good accuracy through the calculation of lifetime ratios.  While the agreement of the ratios between experimental measurements and HQE is excellent for $B$ mesons~\cite{ratios}, 
there are remaining discrepancies between experimental results and theoretical predictions for $b$ baryons.
Recently, the CDF Collaboration~\cite{CDFlifetime2011} used the exclusive decay \Lbch~to report the single most precise determination of the \Lb~lifetime which is more than $2$~standard deviations higher than the world average~\cite{PDG} and slightly higher than the \B0~lifetime. The CDF measurement of the lifetime ratio, $\tau(\Lambda_b^0)/\tau(B^0)$, is higher than the HQE calculation including $\mathcal{O}(1/m_b^4)$ effects, $ 0.88 \pm 0.05$~\cite{lifetimetheory2006}. On the other hand, theoretical predictions are in 
agreement with measurements by the \D0~Collaboration in the \jpsi\La0~\cite{D0lifetime2007} and semileptonic~\cite{D0lifetime2007semileptonic} channels, by the CDF Collaboration in the $\Lambda_c^+ \pi^-$ final state~\cite{CDFlifetime-LbtoLcp-2010},  by the DELPHI, OPAL and ALEPH Collaborations in semileptonic decays~\cite{Delphi, Opal, Aleph}, and previous measurements also in semileptonic channels by the CDF Collaboration~\cite{CDFRunI}.
More  measurements of the \Lb~lifetime and of the ratio $\tau(\Lambda_b^0)/\tau(B^0)$ are required to resolve this discrepancy.

In this article we report a measurement of the \Lb~lifetime using the exclusive decay \Lbch.  
The \B0~lifetime is also measured in the topologically similar channel \Bch. This provides a cross-check of the measurement procedure, and allows the lifetime ratio to be determined directly.
The data used in this analysis were collected with the \D0~detector during the complete Run II
of the Tevatron Collider, from 2002 to 2011, and correspond to an integrated luminosity of 10.4~\ifb~of \ppbar~collisions at a center of mass energy $\sqrt{s} = 1.96$~TeV.  

 A detailed description of the D0 detector can be found in Refs.~\cite{d0det,L1cal, Layer0,SMT}.
 Here, we describe briefly the most relevant detector components used in this analysis. The \D0 central tracking system is composed of a silicon microstrip tracker (SMT) and a central scintillating fiber tracker (CFT)
 immersed in a 2~T solenoidal field.
 The SMT and the CFT are optimized for tracking and vertexing for the pseudorapidity region $|\eta|<3.0$ and $|\eta|<2.0$, respectively, where $\eta \equiv -\ln[\tan(\theta/2)]$ and $\theta$ is the polar angle with respect to the proton beam direction.  Preshower detectors and electromagnetic and hadronic calorimeters surround the tracker.  
A muon spectrometer is located beyond the calorimeter, and consists of three layers of drift tubes and scintillation trigger counters covering $|\eta| < 2.0$. A 1.8~T toroidal iron magnet is located outside the innermost layer of the muon detector.

For all Monte Carlo (MC) simulations in this article, 
we use {\sc pythia}~\cite{Pythia}  to simulate the \ppbar~collisions, {\sc evtgen}~\cite{EvtGen} for modeling the decay of particles containing
$b$ and $c$ quarks, and {\sc geant}~\cite{Geant} to model the detector response.
Multiple \ppbar~interactions are modeled by overlaying hits from random bunch crossings onto the MC.

In order to reconstruct the \Lb~and \B0~candidates, we start by searching for $J/\psi \rightarrow \mu^+ \mu^-$ candidates, which are collected by single muon and dimuon triggers. 
The triggers used do not rely on 
the displacement of tracks
from the interaction point.
At least one \ppbar~interaction vertex (PV) must be identified in each event.  
The interaction vertices are found by minimizing a $\chi^2$ function that depends on all reconstructed tracks in the event and uses 
the transverse beam position averaged over multiple beam crossings.
The resolution of the PV 
 is  $\approx 20$~$\mu$m in the plane perpendicular to the beam (transverse plane). 
Muon candidates are reconstructed from tracks formed  by hits in the central tracking system and with transverse momentum ($p_T$) greater than 1~\gevc.
At least one muon candidate in the event must have hits in the inner layer, and in at least one outer layer of the muon detector.
A second muon candidate, with opposite charge,
must either be detected in the innermost layer of the muon system
or have a calorimeter energy deposit consistent with that of a minimum-ionizing particle along the direction of hits extrapolated from the central tracking system. Each muon track is required to have at least 2 hits in the SMT and 2 hits in the CFT to ensure a high quality common vertex.  
The probability associated with the vertex fit must exceed 1\%.
The dimuon invariant mass is required to be in the range $2.80  - 3.35$~\gevcc, consistent with the \jpsi~mass.

Events with \jpsi~candidates are reprocessed with a version of the track reconstruction algorithm that increases the efficiency for tracks with low 
$p_T$
and high impact parameter~\cite{CasacadeB}. 
We then search for $\Lambda^0 \rightarrow p\pi^-$ candidates reconstructed from pairs of oppositely charged tracks.
The tracks must form a vertex  with a probability associated with the vertex fit greater than 1\%.
The transverse impact parameter significance (the transverse impact parameter with respect to the PV divided by its uncertainty) 
for the two tracks forming \La0~candidates must exceed 2,
and 4 for at least one of them. Each \La0~candidate is required to have a mass in the range
$1.105 - 1.127$~\gevcc. 
The track with the higher $p_T$ is assigned the proton mass. MC simulations indicate that this is always the correct assumption, given the track $p_T$ detection threshold of 120~MeV/$c$.
To suppress contamination from decays of more massive baryons such as $\Sigma^0 \rightarrow \Lambda^0 \gamma$ and $\Xi^{0} \rightarrow \Lambda^0 \pi^{0}$,  
the \La0~momentum vector must point within 1 degree back to the \jpsi~vertex.
The same selection criteria are applied in the selection of $K_S^0 \rightarrow \pi^+\pi^-$ candidates, except that the mass window is chosen in the range  
$0.470 - 0.525$~\gevcc~and pion mass assignments are used. Track pairs simultaneously reconstructed as both \La0~and \K0, due to different mass assignments to the same tracks, are discarded from both samples. This requirement rejects 23\% (6\%) of the \Lbch~(\Bch) signal, as estimated from MC,
without introducing biases in the lifetime measurement.
The fraction of background rejected by this requirement is  58\% (48\%) 
as estimated from data. 
It is important to remove these backgrounds from the samples to avoid
the introduction of biases in the lifetime measurements.

The \Lb~candidates are reconstructed by performing a kinematic fit that constrains the dimuon invariant mass to the world-average \jpsi~mass~\cite{PDG}, and the \La0~and two muon tracks to a common vertex, where the \La0~has been extrapolated from its decay vertex according to the reconstructed \La0~momentum vector. The invariant mass of the \Lb~candidate is required to be within the range $5.15 - 6.05$~\gevcc. The PV is recalculated excluding the \Lb~final decay products. 
The final selection requirements are obtained by maximizing $\mathcal{S}  = S/\sqrt{S+B}$, where $S$ ($B$) is the number of signal (background) candidates in the data sample: the decay length of the \La0~(measured from the \Lb~vertex) and its significance are required to be greater than 0.3~cm and 3.5, respectively; the $p_T$ of the \jpsi, \La0~and \La0~daughter tracks are required to be greater 4.5, 1.8 and 0.3~\gevc, respectively; and the isolation of the \Lb~\footnote{Isolation is defined as  $p(B)/\left[ p(B) + \sum_{<\Delta R}p  \right]$, where $p(B)$ is the  momentum of the $b$ hadron and the sum, excluding the decay products of the $b$~hadron, is over the momentum of all particles from the PV within the larger $\Delta R(\mu^{\pm},b~\text{hadron})$ cone in  pseudorapidity-azimuthal angle space, defined as  $\Delta R = \sqrt{\Delta\eta^2+\Delta\phi^2}$.} is required to be greater than 0.35.
After this optimization, if more than one candidate is found in the event,
which happens in less than 0.3\% of the selected events,
the candidate with the best \Lb~decay vertex fit probability is chosen.
We have verified that this selection is unbiased by varying the selection values chosen by the optimization as described in more detail later.
The same selection criteria are applied to \Bch~decays, except that the $B^0$ mass window is chosen in the range $4.9 - 5.7$~\gevcc.

\begin{figure*}
\includegraphics[scale=0.445]{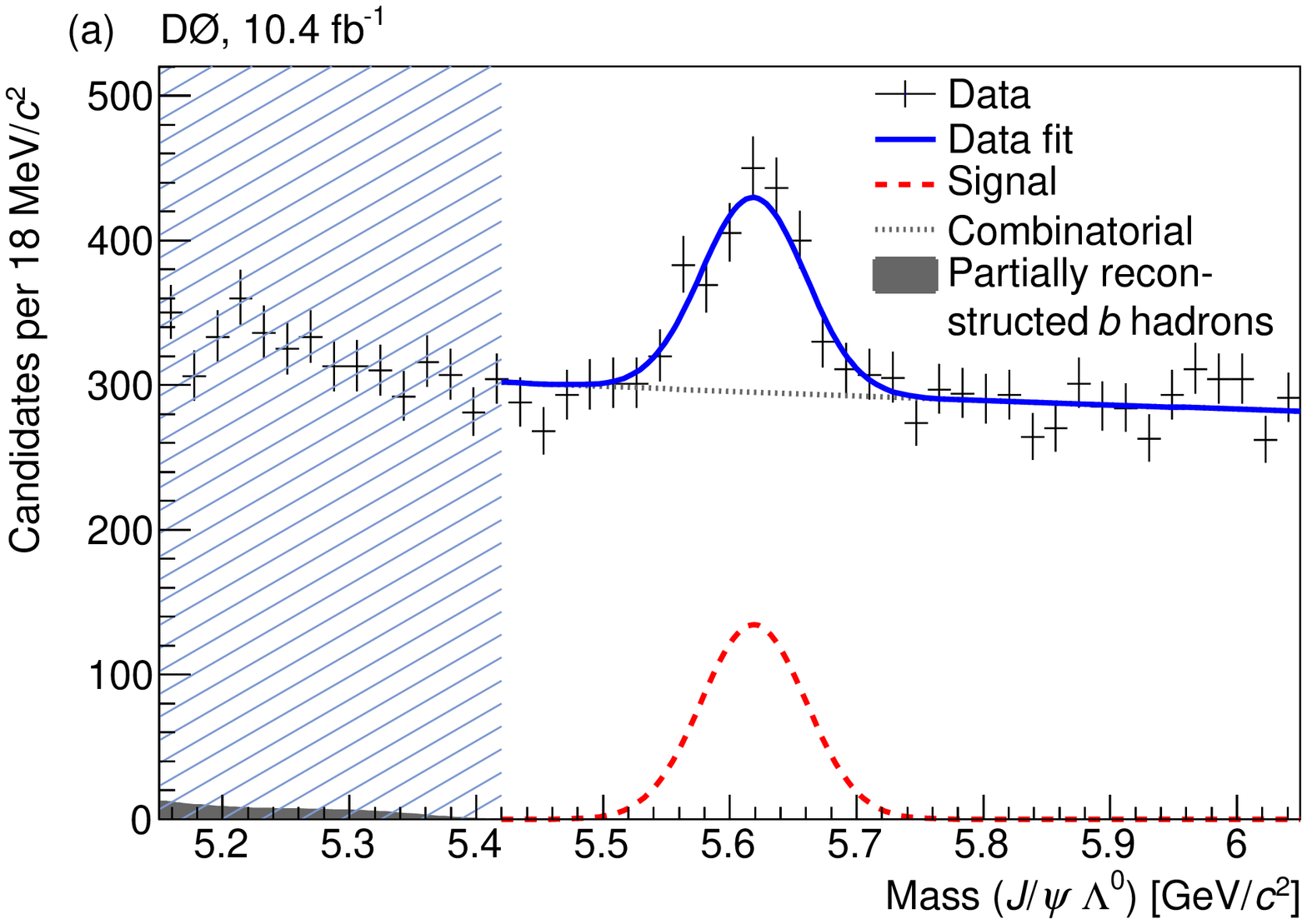} 
\includegraphics[scale=0.445]{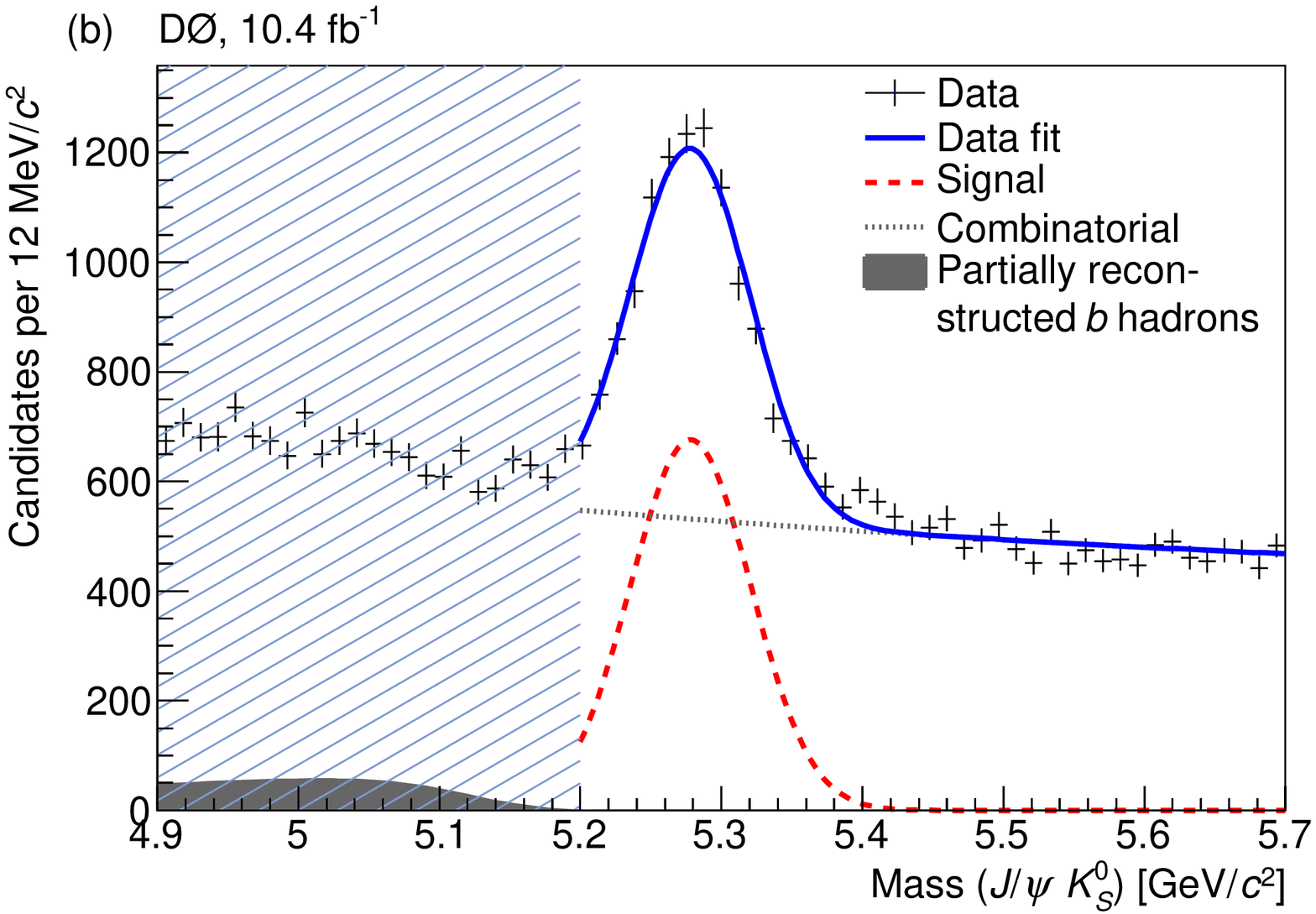}
\caption{\label{fig:massctcut}  (color online) Invariant mass distributions for (a) \Lbch~and (b) \Bch~candidates, with fit results superimposed.
Events in mass regions contaminated with partially reconstructed $b$ hadrons (hatched region) are excluded from the maximum likelihood function used to determine the \Lb~and \B0~lifetimes.}
\end{figure*}

The samples of \Lb~and \B0~candidates have two primary background contributions: combinatorial background and partially reconstructed $b$ hadron decays. 
The combinatorial background can be divided in two categories: prompt background, which accounts for $\approx 70\%$ of the total background, primarily due to direct production of \jpsi~mesons; and non-prompt background, mainly produced by random combinations of  a \jpsi~meson from a $b$ hadron and a \La0~(\K0) candidate  in the event. 
Contamination from  partially reconstructed $b$ hadrons  come from $b$ baryons ($B$ mesons) decaying to a \jpsi~meson, a \La0~baryon (\K0~meson), and additional decay products that are not reconstructed. 

We define the transverse proper decay length as $\lambda = cM L_{xy}/p_T$, 
where $M$ is 
the mass of the $b$ hadron taken from the PDG~\cite{PDG}, and $L_{xy}$ is the vector pointing from the PV to the $b$ hadron decay vertex projected on the $b$ hadron transverse momentum ($\vec{p}_T)$ direction. Due to the fact that signal and partially reconstructed $b$ hadron decays have similar $\lambda$ distributions that are particularly hard to disentangle in the lifetime fit, we remove partially reconstructed $b$ hadrons by rejecting events with \Lb~(\B0) invariant mass below 5.42 (5.20)~\gevcc~from the \Lb~(\B0) sample, as shown in Fig.~\ref{fig:massctcut}. This figure shows the \Lb~and \B0~invariant mass distributions with results of  unbinned maximum likelihood fits superimposed,  excluding events in zones contaminated by partially reconstructed $b$ hadrons. The signal peak is modeled by a Gaussian function. The combinatorial background is parametrized by an exponentially decaying function, while partially reconstructed $b$ hadrons are derived from MC. 
 It can be seen from Fig.~\ref{fig:massctcut} that partially reconstructed $b$ hadrons contribute minimally to the  signal mass region.

In order to extract the lifetimes, we perform separate unbinned maximum likelihood fits for \Lb~and \B0~candidates. The likelihood function ($\mathcal{L}$) depends on 
the probability of reconstructing each candidate event $j$ in the sample with the mass $m_j$, the proper decay length $\lambda_j$ and proper decay length uncertainty $\sigma^{\lambda}_j$:
\begin{equation}\label{eq:likelihood}
\mathcal{L} = \! \prod_{j}  \left [ f_s\mathcal{F}_s(m_j,\lambda_j,\sigma^\lambda_j) + (1\!-\!f_s)\mathcal{F}_b(m_j,\lambda_j,\sigma^\lambda_j) \right ]\!,
\end{equation}
where $f_s$ is the fraction of signal events, and  $\mathcal{F}_s$ ($\mathcal{F}_b$) is the product of the probability distribution functions that model each of the three observables being considered for signal (background) events. The background is further divided into prompt and non-prompt components. For the signal, the mass distribution is modeled by a Gaussian function; the $\lambda$ distribution is parametrized by an exponential decay, $e^{-\lambda_j/c\tau}/c\tau$, convoluted with a Gaussian function  $\mathcal{R} =  e^{-\lambda_j^2/2(s\sigma_j^\lambda)^2}/\sqrt{2\pi } s \sigma_j^\lambda$ that models the detector resolution;  the $\sigma^\lambda$ distribution is obtained from MC simulation and parametrized by a superposition of Gaussian functions.
Here $\tau$ is the lifetime of the $b$ hadron, and the event-by-event uncertainty $\sigma_j^\lambda$ is scaled by a global factor $s$ to take into account a possible underestimation of the uncertainty. For the background, the mass distribution of the prompt (non-prompt) component is modeled by a constant (exponential) function as observed in data when the requirement  $\lambda >100$~$\mu$m  is imposed; the prompt component of the $\lambda$ distribution is parametrized by the resolution function, and the non-prompt component by the 
superposition of two exponential decays for $\lambda < 0$ and two exponential decays for $\lambda> 0$,
as observed from events in the high-mass sideband 
of the $b$ hadron peak  (above 5.80  and 5.45~\gevcc~for \Lb~and \B0, respectively). Finally, the background $\sigma^\lambda$ distribution is modeled by two exponential functions convoluted with a Gaussian function as determined empirically from the high-mass sideband region.
In total, there are 19 parameters in each likelihood fit: lifetime, mean and width of the signal mass,  signal fraction, prompt background fraction, one non-prompt background mass parameter, 7 non-prompt background $\lambda$ parameters, 5 background $\sigma^\lambda$ parameters, and one resolution scale factor.

The maximum likelihood fits to the data yield  
$c\tau(\Lambda_b^0) = 390.7 \pm 22.4$~$\mu$m 
and 
$c\tau(B^0)=   452.2 \pm  7.6$~$\mu$m. 
The numbers of signal events, derived from $f_s$, are $755 \pm 49$ (\Lb) and $5671 \pm 126$ (\B0). 
Figure~\ref{fig:ct}  shows the $\lambda$ distributions for the \Lb~and the \B0~candidates. Fit results are superimposed.

\begin{figure*}
\includegraphics[scale=0.445]{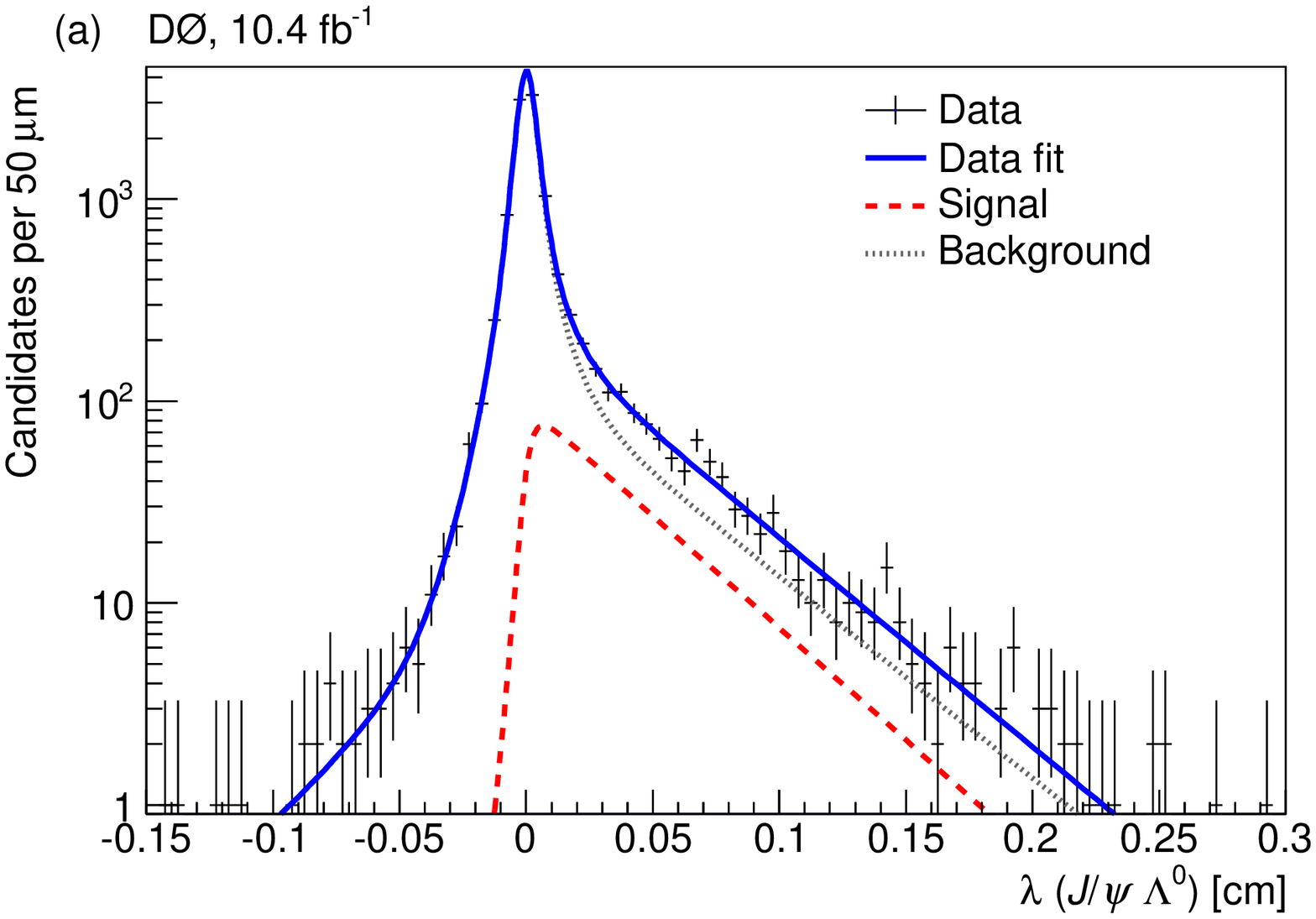} 
\includegraphics[scale=0.445]{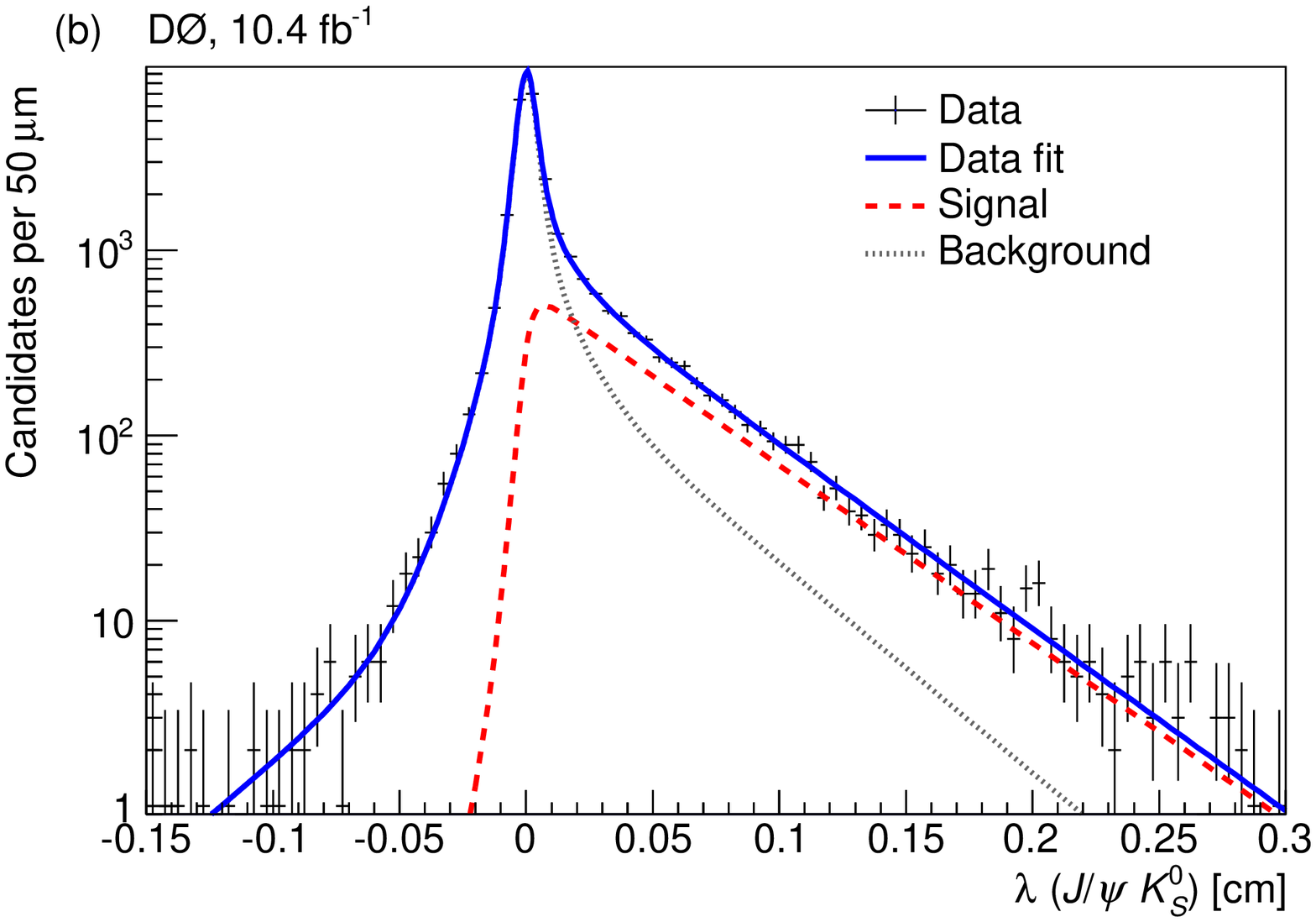}
\caption{\label{fig:ct}  (color online) Proper decay length distributions for (a) \Lbch~and (b) \Bch~candidates, with fit results superimposed.}
\end{figure*}

We investigate possible sources of systematic uncertainties on the measured lifetimes related to  
 the models used to describe the mass, $\lambda$, and $\sigma^\lambda$ distributions.  For the mass we consider
a double Gaussian to model the signal peak instead of the nominal single Gaussian, an exponential function for the prompt background in place of a constant function,
and a second order polynomial for the non-prompt background. The alternative mass models are combined in a single maximum likelihood fit to take into account correlations between the effects of the different models, and the difference with respect to the result of the nominal fit is quoted as the systematic uncertainty
on the mass model.
For $\lambda$ we study the following variations: the introduction of a second Gaussian function along with a second scale factor to model the resolution, the exponential functions in the non-prompt background  replaced by exponentials convoluted with the resolution function, one non-prompt negative exponential instead of two, and one long positive exponential together with a double-Gaussian resolution as a substitute for two non-prompt exponentials and one Gaussian resolution.
All $\lambda$ model changes are combined in a fit, and the difference between the results of this fit and the nominal fit is quoted as the systematic uncertainty due to $\lambda$ parametrization. For $\sigma^\lambda$ we use two different approaches: we use the 
distribution extracted from data by background subtraction, 
parameterized similarly to the nominal background $\sigma^\lambda$ model, instead of the MC model,
and
we use $\sigma^\lambda$ distributions from MC samples generated with different \Lb~(\B0) lifetimes.
The largest variation in the lifetime (with respect to the nominal measurement) between these two alternative approaches is quoted as the systematic uncertainty due to $\sigma^\lambda$ parametrization.  
Residual effects due to contamination from partially reconstructed $b$ hadrons in the samples
are investigated by 
changing the requirement on the invariant mass of the \Lb~and \B0~candidates which are included
in the likelihood fits: the threshold is moved to lower (higher) invariant masses by 40~(20)~MeV/$c^2$, where  40~MeV/$c^2$ is
the resolution on the invariant mass of the reconstructed signal.
The largest variation in the lifetime is quoted as the systematic uncertainty due to possible contamination from partially reconstructed $b$ hadrons. 
In the lifetime fit the contamination from the fully reconstructed decay $B_s^0 \rightarrow J/\psi K_S^0$ is assumed to have little impact on the final result. To test this assumption 
the  $B_s^0 \rightarrow J/\psi K_S^0$ contribution is included in the non-prompt component.
The lifetime shift is found to be negligible.
%
%
The systematic uncertainty due to the alignment of the SMT detector was estimated
in a previous study~\cite{D0lifetime2007} by reconstructing the \B0~sample with the positions of the SMT sensors shifted outwards radially by the alignment uncertainty
and then fitting for the lifetime. 
The systematic uncertainties are summarized in Table~\ref{tab:syst}.

\begin{table}
\caption{\label{tab:syst}Summary of systematic uncertainties on the measurements of $c\tau(\Lambda_b^0)$ and $c\tau(B^0)$, and on their ratio. Individual uncertainties are combined in quadrature to obtain the total uncertainties.  }
\begin{ruledtabular}
\begin{tabular}{lccc}
Source & \Lb~($\mu$m) & \B0~($\mu$m) & Ratio \\
\hline
Mass model & 2.2  &   6.4 &	  0.008 \\
Proper decay length model   &  7.8 &	3.7 &    0.024 \\
Proper decay length uncertainty  	& 2.5	& 8.9  &	 0.020 \\
Partially reconstructed $b$ hadrons	&  2.7	& 1.3  &	0.008 \\
$B_s^0 \rightarrow J/\psi K_S^0$	& --	& 0.4	&    0.001 \\
Alignment	&  5.4	&  5.4	 &  0.002 \\\hline
Total  &	 10.4 	 & 12.9 	&	0.033 \\
\end{tabular}
\end{ruledtabular}
\end{table}

We perform several cross-checks of the lifetime measurements. 
We extract the signal yield in bins of $\lambda$ by fitting the mass distribution in each of these regions. From these measurements, lifetimes are obtained by the $\chi^2$ minimization of the signal yield expected in each $\lambda$ bin according to the first term in Eq.~\ref{eq:likelihood}. While this method is statistically inferior with respect to the maximum likelihood fit, it is also less dependent on the modeling of the different background components. The results of this study are $c\tau_{\Lambda_b^0} = 391.4 \pm 35.8\text{ (stat.)}~\mu$m and $c\tau_{B^0} = 458.3 \pm 8.9 \text{ (stat.)}~\mu$m. 
%
The sample is also split into different data taking periods, $\eta$ regions, and numbers of hits in the SMT detector. 
All results obtained with these variations are consistent with our measurement.
In order to check that the optimization procedure does not give a potential bias to the selection, we verify that our results remain stable when all requirements in variables used in the optimization process are removed one at a time, when looser and tighter requirements are applied to kinematic variables, and when multiple candidates that pass all selection requirements per event are allowed. 
The results also remain stable after removing the high-end tail 
(above $100$~$\mu$m) of the $\sigma^\lambda$  distribution, mainly populated by background events.
We also cross check the fitting procedure and selection criteria by measuring the \Lb~and \B0~lifetimes in MC events. The lifetimes obtained are consistent with the input values.

In summary, using the full data sample collected by the \D0~experiment, we measure the lifetime of the \Lb~baryon in the \jpsi\La0~final state to be 
\begin{linenomath*}
\begin{equation}
\tau(\Lambda_b^0) =   1.303 \pm 0.075 \text{ (stat.)} \pm 0.035 \text{ (syst.)} \text{ ps} ,
\end{equation}
\end{linenomath*}
consistent with the world-average,  $1.425 \pm 0.032$ ps~\cite{PDG}. The method to measure the \Lb~lifetime is
also used for
\Bch~decays, 
for which we obtain
\begin{linenomath*}
\begin{equation}
\tau(B^0) =   1.508 \pm 0.025 \text{ (stat.)} \pm 0.043 \text{ (syst.)} \text{ ps} ,
\end{equation} 
\end{linenomath*}
in good agreement with the world average, $1.519 \pm 0.007 $~ps~\cite{PDG}.

Using these measurements we  calculate the ratio of lifetimes,
\begin{linenomath*}
\begin{equation}\label{eq:ratioresult}
\frac{\tau(\Lambda_b^0)}{\tau(B^0)}=0.864 \pm 0.052 \text{ (stat.)} \pm 0.033 \text{ (syst.)},
\end{equation}
\end{linenomath*}
where the systematic uncertainty is determined
from the differences between the lifetime ratio obtained for each systematic variation and the ratio of the nominal measurements, and combining theses differences in quadrature,
as shown in Table~\ref{tab:syst}. 
Our result, $0.86 \pm 0.06$, is in good agreement with the HQE prediction of  $0.88 \pm 0.05$~\cite{lifetimetheory2006} and compatible with the current world-average, $1.00 \pm 0.06$~\cite{PDG}, but differs with the latest measurement of the CDF Collaboration, $1.02 \pm 0.03$~\cite{CDFlifetime2011}, at the $2.2$~standard deviations level.
Our measurements supersede the previous \D0~results of $\tau(\Lambda_b^0)$, $\tau(B^0)$  and $\tau(\Lambda_b^0)/\tau(B^0)$~\cite{D0lifetime2007}.

%
We thank the staffs at Fermilab and collaborating institutions,
and acknowledge support from the
DOE and NSF (USA);
CEA and CNRS/IN2P3 (France);
MON, Rosatom and RFBR (Russia);
CNPq, FAPERJ, FAPESP and FUNDUNESP (Brazil);
DAE and DST (India);
Colciencias (Colombia);
CONACyT (Mexico);
NRF (Korea);
FOM (The Netherlands);
STFC and the Royal Society (United Kingdom);
MSMT and GACR (Czech Republic);
BMBF and DFG (Germany);
SFI (Ireland);
The Swedish Research Council (Sweden);
and
CAS and CNSF (China).

\end{document}
%